\documentclass[conference]{IEEEtran}
\makeatletter
\def\ps@headings{%
\def\@oddhead{\mbox{}\scriptsize\rightmark \hfil \thepage}%
\def\@evenhead{\scriptsize\thepage \hfil \leftmark\mbox{}}%
\def\@oddfoot{}%
\def\@evenfoot{}}
\makeatother
\usepackage{cite}

\ifCLASSINFOpdf
   \usepackage[pdftex]{graphicx}
   \graphicspath{{fig-pdf/}{fig-jpeg/}{fig-png/}}
   \DeclareGraphicsExtensions{.pdf,.jpeg,.png}
\else
   \usepackage[dvips]{graphicx}
   \graphicspath{{fig-eps/}}
   \DeclareGraphicsExtensions{.eps}
\fi
\usepackage[tight,footnotesize]{subfigure}
\usepackage[cmex10]{amsmath}
\usepackage{mdwmath}
\usepackage{mdwtab}

\usepackage{algorithmic}
\usepackage{algorithm}
\floatname{algorithm}{Algorithm}

\usepackage{array}

\usepackage{url}

\usepackage{glossaries}
\newacronym{sns}{SNS}{Social Networking Services}
\newacronym{osn}{OSN}{Online Social Networks}
\newacronym{dsn}{DSN}{Decentralized Social Network}
\newacronym{msn}{MSN}{Mobile Social Network}
\newacronym{dtn}{DTN}{Delay Tolerant Network}
\newacronym{nfc}{NFC}{Near-Field Communication}
\newacronym{d2d}{D2D}{Device-to-Device}
\newacronym{ppr}{PPR}{Personalized PageRank}
\newacronym{pr}{PR}{PageRank}
\newacronym{xmpp}{XMPP}{Extensible Messaging and Presence Protocol}
\newacronym{roc}{ROC}{Receiver Operating Characteristics}
\newacronym{auc}{AUC}{Area Under the ROC Curve}
\newacronym{tpr}{TPR}{True Positive Rate}
\newacronym{fpr}{FPR}{False Positive Rate}
\newacronym{ev}{EV}{Escape Vector}
\newacronym{rw}{RW}{Random Walker}

\newcommand{\req}[1]{{Eq.~{#1}}}
\newcommand{\rfig}[1]{{Fig.~{#1}}}
\newcommand{\rtbl}[1]{{Table~{#1}}}

\newcommand{\rsec}[1]{{Section~{#1}}}
\newcommand{\ralg}[1]{{Algorithm~{#1}}}

\DeclareMathOperator*{\maximize}{maximize}

\newcommand{\tran}[1]{#1^\mathrm{T}}

\usepackage{amsthm}

\newtheorem{myre}{Remark}

\title{Localized Algorithm of Community Detection on Large-Scale 
Decentralized Social Networks}

\author{{\bf  Pili Hu~~~~~~~~~~~~~~~~~~~~~~~~~~~~~~~~Wing Cheong Lau
}\\
Department of Information Engineering,  The Chinese University of Hong Kong, Hong Kong
}

\begin{document}

\maketitle

\begin{abstract}
Despite the overwhelming success of the existing Social Networking Services (SNS), 
their centralized ownership and control have led to serious concerns in user privacy,
censorship vulnerability and operational robustness of these services. To overcome 
these limitations, Distributed Social Networks (DSN) have recently been proposed 
and implemented. Under these new DSN architectures, no single party possesses 
the full knowledge of the entire social network. While this approach solves 
the above problems, the lack of global knowledge for the DSN nodes makes it 
much more challenging to support some common but critical SNS services like 
friends discovery and community detection. In this paper, we tackle the problem 
of community detection for a given user under the constraint of limited local 
topology information as imposed by common DSN architectures. By considering 
the Personalized Page Rank (PPR) approach as an ink spilling process, we 
justify its applicability for decentralized community detection using limited 
local topology information. Our proposed PPR-based solution has a wide range 
of applications such as friends recommendation, targeted advertisement, 
automated social relationship labeling and sybil defense. 
Using data collected from a large-scale SNS in practice, 
we demonstrate our adapted version of PPR can significantly outperform 
the basic PR as well as two other commonly used heuristics. 
The inclusion of a few manually labeled friends in the Escape Vector (EV) can 
boost the performance considerably (64.97\% relative improvement 
in terms of Area Under the ROC Curve (AUC)). 
\end{abstract}

\IEEEpeerreviewmaketitle

\section{Introduction}
\label{sec:intro}

In the past few years, large-scale 
\gls{sns} such as 
Facebook, Twitter and Renren have become a major part of people's daily life.
Besides serving as a  communication and information-sharing platform for their users,  
these  
\gls{osn}
also play a key role in friends discovery and community 
formation for users of common interests.
Despite the overwhelming success of the existing SNS, the centralized ownership and control of these services
have led to serious concerns in user privacy, censorship and  operational robustness.
Since the operator of an SNS has full knowledge of the profiles, social relationships and communication
activities of their users, it is a high-value and obvious target for not only the typical attackers but also many
totalitarian regimes which constantly seek to monitor and control information dissemination among their people. 

To overcome the drawbacks of centralized OSNs, Decentralized Social Networks (DSN) such as Diaspora\cite{prj-diaspora}, Musubi\cite{dodson2012musubi} and 
OneSocialWeb\cite{prj-onesocialweb} have recently been proposed and implemented. 
Under these DSN architectures, no single party possesses the full knowledge of the entire social network. 
For example, in the Diaspora network,  a server (super-node) only has its own partial view of those users registered within it. 
It is even more restrictive in the Musubi network in which
every user only has visibility of his/her own relationships in form of a contact book; even if a user can consult his direct friends 
for their friend lists, his knowledge of the network  is still constrained to 2 hops.  
Current efforts of \gls{dsn} are mainly on system prototyping. 
Although the design of many building blocks for DSN can be 
adapted from prior  P2P system designs, the nature of \gls{dsn} 
imposes further challenges:
Due to privacy and trust concern, 
available information for a single node is limited. 
For example, in a BT swarm, nodes are willing to 
exchange content with complete strangers 
as long as they provide each other enough upload rate. 
In \gls{dsn}, however, people only want to share contents
(profile, status, blog, album, etc) within certain scope.

While the compartmentalized approach of the DSNs can enhance user privacy and make them less vulnerable to
monitoring, censorship and sabotage attacks,  the lack of global knowledge for the DSN nodes makes it much more 
difficult to support some common but critical services such as friend discovery/ recommendations and community detection.
In particular, the DSN architecture invalidates  many  assumptions on data availability and global network topology awareness 
which are required by most existing community detection algorithms. 
To address this challenge, this paper considers the following problem for the \gls{dsn}s:
Given a user and his/her partial knowledge 
of the social graph, can we predict which nodes are in the same {\it community} with him?
The solution of this problem can support 
many fundamental services of \gls{dsn}
including friends recommendation, targeted advertisement, 
automated relationship labeling and sybil defense. 
In particular, we will focus on the design and performance evaluation of
{\it localized} and {\it fully-decentralized} community detection algorithms for {\it large-scale} \gls{dsn}s based on the Personalized Page Rank (PPR) approach:
\begin{itemize}
	\item By ``localized'', 
		we mean that every node's knowledge of the network is 
		limited to its local neighborhood topology
		\footnote{While it has been shown that one can substantially improve community detection performance by incorporating information beyond mere topological data \cite{sachan2012using,lin2012community}, as an initial study of community detection on \gls{dsn}, we consider only the topological data of the network for now.}. 
		This is in contrast to other so-called localized approaches, 
		e.g. the local graph partitioning algorithm in \cite{andersen2006local},
		for which global network topology is actually available even though 
		the algorithm may choose not to fully explore or leverage such information.

	\item By ``fully-decentralized'', 
		we mean that every individual node within the DSN can execute 
		our proposed community detection algorithm based on its locally acquired knowledge. 
		Explicit coordination from other nodes is not needed. 
		In contrast,  many existing  ``distributed community detection'' algorithms, 
		e.g. those proposed in  \cite{hui2007distributed} and \cite{ramaswamy2005distributed}, 
		do require exchange of information and collaboration among the network nodes.

	\item By ``large-scale'', 
		we mean that our algorithm is designed to be scalable for DSNs of sizes comparable to the existing top-tier OSNs. 
		In fact,  we evaluate our proposed algorithm in \rsec{\ref{sec:data}} using
		real data collected from the Renren social network in China (which already had 160 million registered users by 2011\cite{wiki_renren}). 
		More importantly, even if we only consider the 2-hop local neighborhood of a target Renren user, 
		the size of the topological data is already very substantial: 
		on average, each user has around 350 direct friends and 75,000 friends-of-friends. 
		In fact, the size of the local topology of a single target user in our study is already 
		as large as many medium-sized data sets used for evaluating many 
		global community detection algorithms reported in the literature.
\end{itemize}

In short, this paper has made the following technical contributions:
\begin{itemize}
	\item We formulate the new problem of community detection 
	under the constraint of limited local topology awareness found in DSNs.

         \item We adapt the Personalized Page Rank (\gls{ppr}) algorithm for community detection under limited topology information.
         In particular, by interpreting  \gls{ppr} as an ink spilling process, we justify its applicability for community detection for a target user.
   
	\item We investigate different design choices of the Escape Vector (EV) for \gls{ppr}.  Their impact on community detection performance are evaluated using real-world data sets collected from the large-scale Renren social network.

\end{itemize}

The rest of the paper is organized as follows. 
In \rsec{\ref{sec:rwork}}, we survey related works. 
In \rsec{\ref{sec:form}}, we formulate the community 
detection problem on 2-hop-only topology of the 
social network graph. 
In \rsec{\ref{sec:algorithm}}, we analyze the problem 
and propose our algorithm based on the \gls{ppr} approach. 
In \rsec{\ref{sec:eval}}, we evaluate the performance of our \gls{ppr}
proposal by comparing it with other commonly-deployed heuristics, 
and study the effect of different choices of \gls{ppr} Escape Vectors in depth. 
We then conclude the this paper in \rsec{\ref{sec:conc}}. 

\section{Related Work}
\label{sec:rwork}

\subsection{Community Detection Algorithms}

Community detection is a classical problem of 
partitioning a graph into multiple (possibly overlapping) subsets (communities)
while satisfying the following conditions: 
intra-community linkage is dense and
inter-community linkage is sparse.
In the literature, this principle has been realized in 
many different ways including 
Modularity maximization, 
Conductance minimization and 
Normalized Cut minimization
\cite{aggarwal2011social}. 
So far, Modularity maximization \cite{newman2004-finding} has been the most popular 
approach for community detection. However, since the resultant combinatorial optimization problem is NP-Hard, 
researchers have proposed many heuristics to obtain approximate solutions. 
Newman \cite{newman2006-Finding} proposed an eigen decomposition approach plus 
a local search algorithm.
Agarwal \cite{agarwal2008modularity} leveraged several 
mathematical programing techniques. 
Those works are the first batch that focus on 
detecting community based on only full topology information. 
 
Since pure topology-based community detection algorithms are well developed by now,
many researchers are trying to incorporate more side information
to enhance performance of existing algorithms. 
In \cite{sachan2012using}, Sachan et al combine 
social graph topology, interaction pattern, 
and topics to discover topically meaningful communities. 
In \cite{lin2012community}, Lin studies the scenario 
where node-level information is complete
(like authors' bag-of-words coordinates) 
but only part of the topology is observed. 
Note however that complete node-level information 
is still too hard to be acquired in the \gls{dsn} scenario.

All of the above community detection algorithms are 
centralized ones requiring global information. 
Besides the data constraint, those global algorithms 
are usually computational intensive, 
which makes them inapplicable for  
large-scale OSNs.  As such, our study of scalable and localized algorithms 
is important even under the centralized \gls{osn} settings. 

It is worth to note that researchers have proposed some algorithms 
to tackle with large-scale graphs in centralized settings. 
\cite{raghavan2007near} proposed a Label Propagation (LP) algorithm
whose complexity for one iteration is linear of number of edges. 
LP initializes nodes with their unique labels and nodes switch to 
the majority label owned by their neighbours at each iteration. 
\cite{leung2009towards} conducted more experiments on the LP algorithm
and proposed several improvements. 
Since our algorithm only requires local topology of each node, 
it is easy to be parallelized and scale to large graphs naturally under the centralized settings. 
We leave the combination of multiple runs of our localized algorithm to 
form a global community to future work. 

\subsection{Friend Recommendation Systems in Practice}

We mentioned several applications of community detection 
algorithms on \gls{dsn} in \rsec{\ref{sec:intro}}. 
In the literature, Friend Recommendation is also 
associated with the problem of link prediction\cite{liben2007link}. 
Although many sophisticated algorithms have been developed, 
they also suffer from their forbidding computational requirement
like global community detection algorithms do. 
For large-scale OSN in practice, simple heuristics are believed to be in use. 
Veneta \cite{von2008veneta}, for example, uses
Common Neighbor to do friend discovery but their
focus is on how to achieve this simple heuristic securely. 
While no detail information  is available for the 
proprietary  friend recommendation algorithms used by  commercial  \gls{osn} service providers,  
a recent post \cite{qq_quan} by the Tecent research team
noted the deficiency of using Common Neighbors heuristic for community detection. It also
expressed their concern of  the computational burden of mature global optimization methods. 
\subsection{Other Topics to Be Assisted by Community Detection}

Community detection output can assist many other topics. 
Detecting community in decentralized scenario can also help research in those topics. 
In this section, we briefly discuss a few of them. 

To design the DTN routing algorithm, BubbleRap\cite{hui2008bubble}, 
Pan used the knowledge of community and centrality as heuristics. 

It is shown that video views preserves good locality
in terms of both geo-locations\cite{brodersen2012youtube}
and social network topology\cite{wang2012guiding}. 
The key observation is that, many interest diffusion 
processes have some target communities. 
Community detection result can help to improve the performance of those 
social network assisted content distribution. 

Social network assisted Sybil Defense also draws a lot of attention in recent years. 
It includes sybil identity detection and sybil community detection\cite{yu2011sybiltut}. 
The latter one is the same problem as community detection. 
What differs %
is that the structural assumptions in Sybil Defense schemes are usually stronger
and its application is more 
sensitive to TP or FP depending on the scenario. 
With locally detected communities, nodes can share information to form a trust region, 
thus excluding nodes who are highly susceptible to be sybil. 

\section{Problem Formulation}
\label{sec:form}

\subsection{Define Local Topology Using The Notion of Hops}

We model the social network as a graph
where each node corresponds to a person 
and each link corresponds to a relationship. 
For social networks in practice, relationships can be 
directed (like following on Twitter) 
or undirected (like being friends on Facebook). 
In this work, we focus on undirected relationship only. 

Let's denote the undirected social graph as $ G=(V,E) $, 
where $ V $ is the vertex (node) set and 
$ E $ is the edge (link) set.
We also denote  a node $ v_i \in V $ 
using the shorthand notation $ i $. 
Given a node $ i $, we denote its $0$-hop neighbour set as $N^{(0)}_i = \{i\}$. 
Iteratively, we can define the $h$-hop neighbour set as 
$ N^{(h)}_i = \{ j | k \in N^{(h-1)}_i, j \in V, (k,j) \in E \} \cup N^{(h-1)}_i $,
where $ h \ge 1 $. 
The set of edges that can be observed within $h$-hop is
$ E_i^{(h)} = \{ (j, k) | j \in N^{(h-1)}_i, k \in N^{(h)}_i, (j, k) \in E \}$. 

Using the above notations, we can define the local
topology information available to a user as follows:
We call a normal user of \gls{sns} as ``observer'',  
denoted by $ o $. 
How much information the observer can get depends on the application scenario.
Given an observer $ o $, the $ h $-hop local topology is 
$ G^{(h)}_o = (N^{(h)}_o, E_o^{(h)}) $. 
To facilitate  discussions, we define 
$ n = |N^{(h)}_o| $ 
to be the number of nodes and 
$ m = |E_o^{(h)}| $ 
to be the number of edges.

\subsection{Illustration of 2-Hop Topology}

\begin{figure}[!t]
	\centering
        \subfigure[2 Hop Illustration]{%
            \label{fig:2hop_pic}
            \includegraphics[width=0.49\linewidth]{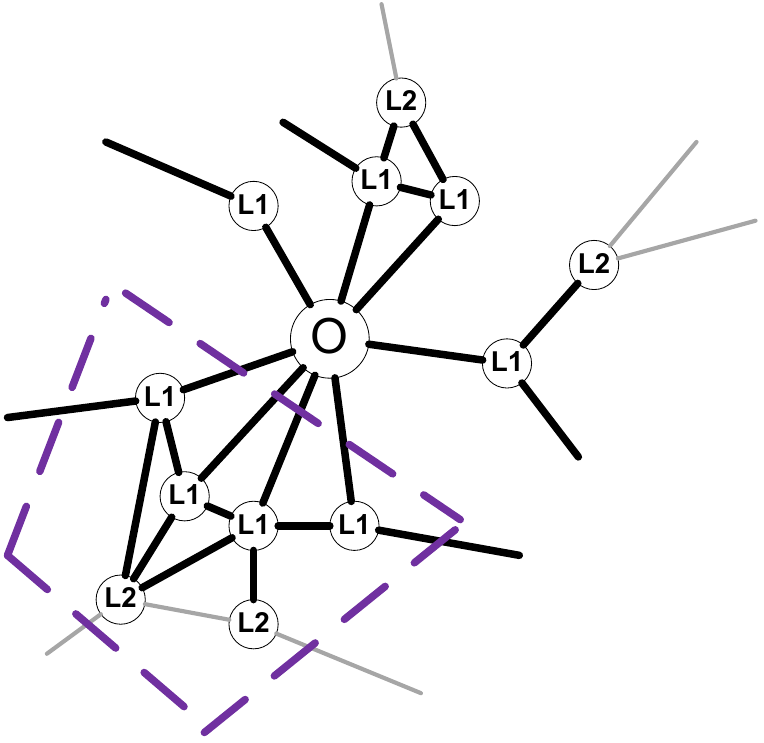}
        }%
        \subfigure[Matrix Illustration]{%
            \label{fig:2hop_matrix}
            \includegraphics[width=0.49\linewidth]{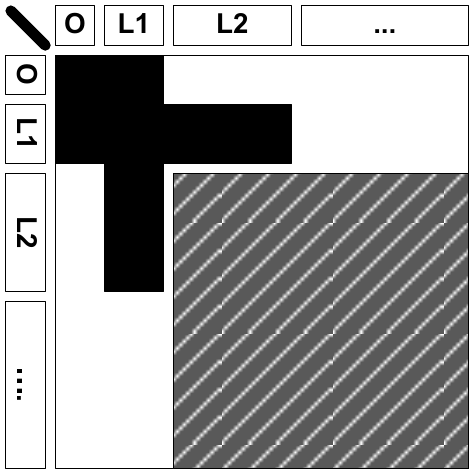}
        }%
    \caption{2 Hop Topology Illustration and Matrix Form}
    \label{fig:2hop}
\end{figure}

\rfig{\ref{fig:2hop}} gives an illustration of $ G^{(2)}_o $. 
In \rfig{\ref{fig:2hop_pic}}, the central 
biggest node(O) is the observer. 
Nodes denoted by ``L1'' are the direct friends of the observer, also 
referred to as the level 1 nodes (reachable within 1 hop). 
Similarly, nodes denoted by ``L2'' can be reached within 2 hops and
are referred to as level 2 nodes. Dark thick edges show the links 
which are visible to the observer.
Light thin edges show
possibly existing links but the observer cannot see them within its 
2 hop visibility. Dashed region shows a probable
community since those nodes have dense linkage inside 
and sparse linkage outside.
\rfig{\ref{fig:2hop_matrix}}
illustrates the observer's knowledge of the adjacency matrix. Nodes
are ordered and labeled using the same convention as the 
left panel. The black 
area is the part known to the observer. The white vacancy represents 
zeros in the adjacency matrix. The shadowed square is unknown, 
including the light thin edges in the left figure.

\subsection{Justifications for a 2-Hop Formulation}

As an initial study of community detection on \gls{dsn}, 
we formulate the problem assuming only 2-hop topology information is available for the target user. 
The reasons are:
\begin{itemize}	
	\item \textbf{2-Hop is the smallest topology to study}. 
	The 1-hop subgraph from an observer is a simple star, 
	and there is no further topology information to utilize. 
	\item \textbf{OSN diameter continues to shrink}. 
	Dating back to 1929, Frigyes Karinthy proposed the concept  
	of 6 degree(7 hops) separation \cite{wiki_chains}. 
	Stanley Milgram conducted experiments in 1960 to show 
	an average degree of 5.2 (6 hops)\cite{fbdata_anatomy}. 
	According to Facebook's study in the end of 2011
	\cite{fbdata_anatomy}, the separation in the Facebook network was 
	5.28 hops in 2008 but became 4.74 hops in 2011.
	\item \textbf{Average node degree is very high}. 
	Dunbar's Number \cite{dunbar2010dunbar} estimates 
	the number of stable relationships of a typical person 
	to be 150. 
	The Renren network
	(further explained in \rsec{\ref{sec:data}}) 
	is much better connected than that and the size of topology data associated with
	a local neighborhood would increase rapidly with the hop-count.
	\item \textbf{There are existing heuristics on only 2-hop topology}. 
	Heuristics like common neighbors and Adamic/Adar
	score\cite{aggarwal2011social}
	(further explained in \rsec{\ref{sec:roc_4f}}) 
	can be computed exactly within 2 hop 
	and provide reasonable baseline for our study.  
	\item \textbf{Trust concern prevents user from 
	looking further away}. 
	In the extreme case of \gls{dsn}, users only have their 
	own first hop topology
	(e.g. through Musubi's contact book). 
	The 2-hop topology can be obtained by requesting those direct friends
	and is reliable. 
	However, asking a stranger for his buddy list 
	may risk at polluting the raw data 
	and resulting in meaningless detection result. 
\end{itemize}

\subsection{Community Detection As a Classification problem}

Conventional community detection is formulated 
as a clustering problem. 
That is, given the full graph $ G=(V,E) $, 
partition the vertex set into $ K $ 
subsets $ \{C_1, C_2, \ldots, C_K\} $, 
such that $ \cap_{i=1}^{K}C_i = \emptyset $
and $ \cup_{i=1}^{K}C_i = V $.
The notion of community is that nodes within 
the same community have dense linkage and
nodes from different communities have sparse linkage. 
To implement this concept mathematically, 
people proposed different quality measures $ Q(.) $
on a partitioning. 
Thus the clustering version of community 
detection problem becomes
(consider the non-overlapping case)
\begin{eqnarray}
	\maximize_{\{C_1, C_2, \ldots, C_K\}} & & Q(\{C_1, C_2, \ldots, C_K\}) \\
	s.t. & & \cap_{i=1}^{K}C_i = \emptyset \\
	s.t. & & \cup_{i=1}^{K}C_i = V
\end{eqnarray}
Whether the ``maximize'' or ``minimize'' operator is used
depends on the nature of the quality measure. 
One implementation of $ Q(.) $ is the Modularity 
proposed by Newman\cite{newman2004-finding}:
\begin{equation}
	Q = \frac{1}{2m}\sum_{k=1}^{K}\sum_{i,j \in C_k} (A_{ij} - \frac{d_id_j}{2m})
\end{equation}
where $ A $ is the adjacency matrix of $ G $, 
$ d_i = \sum_{j}{A_{ij}} $ is the degree of node $ i $, 
$ m = |E| $ is the total number of edges. 
Modularity in essence measures how far the 
resulting partition is from a random graph
and, the higher the better. 
There are other quality measures, e.g. Conductance. 
Interested readers can refer to \cite{aggarwal2011social}
for more information. 

The clustering formulation draws a lot of interest in the past. 
One reason is that the form is clean and highly amenable for theoretical studies. 
The other reason is that at the time when people started
to do community detection, there were no successful 
large-scale \gls{osn}s. 
Researchers did not have ground-truth to 
validate whether a community partitioning is {\it correct}. 
So the clustering formulation and quality function
evaluation became the mainstream approach in the research community. 

With the wide acceptance of \gls{osn} in the recent years, 
labeled data become available. 
For instance, in the Renren social network graph, every node is associated
with an institution name $ I_i $, representing his/her university, 
high school, company, etc. 
Note that a person may have multiple institutions in reality. 
For the moment, we consider the social graph with only 
one institution name $ I_i $ for each node $ i $. 
Details regarding multiple institution names are put in \rsec{\ref{sec:data}}.
Since the ground-truth is available, evaluating 
the community detection result against the crawled 
ground-truth is more reasonable.  
Mathematically speaking, given an observer $ o $
and its 2-hop local topology $ G^{(2)}_o $, 
we want to determine $ \forall i \in T_o $
whether $ I_i=I_o $, 
where $ T_o \in V $ is the set of test nodes. 
$ T_o = N^{(2)}_o$ means that we evaluate the algorithm 
on all nodes within 2 hops, which fits the application of 
friend recommendation. 
$ T_o = N^{(1)}_o$ means that we only evaluate the algorithm 
on only level 1 nodes, which fits the application of 
automated friend categorization. 
Both choices of $ T_o $ will be evaluated in \rsec{\ref{sec:eval}}. 
We then cast this decision problem as a binary-classification problem, 
i.e. assign a label $ L_i $ (the ground-truth) to each node:
\begin{equation}
	L_i = \left\lbrace
	\begin{array}{cc}
	1 & I_i=I_o \\
	0 & \text{otherwise}
	\end{array}
	\right.
\end{equation}
The algorithm should take the 2-hop topology as input and 
output predicted labels $ \hat{L}_i \in \{1, 0\}$. 

The confusion matrix of classification result 
is shown in \rfig{\ref{tbl:confm}}. 
Using the notation in the confusion matrix, 
we have multiple standard ways to evaluate the quality 
of predicted labels $ \hat{L}_i $, such as accuracy:
\begin{equation}
	\text{Accuracy} = {(a + d)}/{(a + b + c + d)}
	\label{eq:def_accuracy}
\end{equation}
true positive rate
\begin{equation}
	\text{TPR} = {a}/{(a + b)}
	\label{eq:def_tpr}
\end{equation} and 
false positive rate
\begin{equation}
	\text{FPR} = {c}/{(c + d)}
	\label{eq:def_fpr}
\end{equation}

\glsreset{roc}
\glsreset{auc}

In \rsec{\ref{sec:eval}}, we will further analyze 
the shortcomings of the above evaluation methods
and propose to use \gls{roc} and \gls{auc} to 
evaluate key steps of the algorithm in depth. 

\begin{figure}[!t]
	\centering
	\begin{tabular}{|c|cc|}
	\hline 
	\# of Samples & $ \hat{L}=1 $ & $ \hat{L}=0 $ \\
	\hline
	$ L=1 $ & $ a $ & $ b $ \\
	$ L=0 $ & $ c $ & $ d $ \\
	\hline
	\end{tabular}
	\caption{Confusion Matrix}
	\label{tbl:confm}
\end{figure}

\section{Analysis and Algorithm Design}
\label{sec:algorithm}

\glsreset{pr}
\glsreset{ppr}

In this section, we analyze the problem and propose 
a two-stage algorithm framework first. 
Then we shift our focus on the first stage, 
i.e. finding a reasonably good ranking function of 
all nodes in $ T_o $. 
Among all kinds of possibilities, we investigate 
the variations of \gls{pr} and \gls{ppr}. 
We show \gls{pr} is not informative in our setting
and justifies the potential of \gls{ppr}
using an ``ink spilling'' process. 
In the last part of this section, we analyze 
the complexity of \gls{ppr} computation using 
both conventional matrix multiplication method 
and ink spilling algorithm. 

\subsection{Two-stage Classification Framework}

There are many ways to construct good classifiers. 
In this work, we investigate a simple framework
as shown in \ralg{\ref{alg:framework}}. 
In the first stage, we compute a ranking function $ f(i) $, 
which assigns scores to all the nodes concerned. 
This score reflects how likely one node is positive
but it does not necessarily be 
strict probability distribution on $ T_o $. 
In the later discussion, we also refer to $ f(i) $ as 
``feature'', ``heuristics'' and ``ranking'' interchangeably. 
We can consider the two metrics defined in \req{\ref{eq:common}} and \req{\ref{eq:adar}}
as two examples of $ f(i) $. 
In the second stage, we compute a proper threshold
and cut the nodes into positive and negative sets
according to this threshold. 

Note that our framework itself may not be optimal. 
For example, we can compute more than one feature, 
and train a probably better classifier. 
As a pilot study of the potential of only 2-hop topology, 
we, however, are more interested in what performance
a single ranking function can achieve in this setting. 
The investigation on the use of multi-dimensional ranking function
is left as future work.

\begin{algorithm}[!t]
	\caption{Community Classification Framework}
	\label{alg:framework}
	\begin{algorithmic}[1]
		\REQUIRE Observer $ o $, 2-hop topology $ G^{(2)}_o$, test set $ T_o $
		\ENSURE Predicted labels: $ \hat{L}_i$,  $ \forall i \in T_o $
		\STATE Compute a ranking function
		$ f(i) $, $ \forall i \in T_o $
		\STATE Compute a proper threshold $ R \leftarrow g(o, G_o, f) $
		\STATE Binary classification by thresholding:
		$$ \hat{L}_i \leftarrow \left\lbrace
		\begin{array}{cc}
			1 & f(i) \ge R \\
			0 & f(i) < R
		\end{array}				
		\right.  $$
	\end{algorithmic}
\end{algorithm}

\subsection{PageRank and the Limitation of Its Basic Form}

We denote the adjacency matrix of a graph by $A$. 
The element of degree vector $ \vec{d} $ is defined as 
$ \vec{d}_i = \sum_{j}{A_{ij}} $. 
The degree matrix is defined as 
$ D = \text{diag}(\vec{d}_1, \vec{d}_2, \ldots, \vec{d}_n) $.
Left multiplying $ A $ by the inverse degree matrix, 
we get the normalized adjacency matrix
(also called the walk matrix):
$W = D^{-1}A$. 
The PR vector $\vec{v}$ is obtained by solving the following 
fixed-point equation: 
\begin{equation}
	\vec{v} = \tran{W}\vec{v}
	\label{eq:pr0}
\end{equation}
The solution of $\vec{v}$ can also be interpreted as 
the stationary probability distribution of a \gls{rw} on the graph. 
In each step, the \gls{rw} uniformly picks an adjacent edge
and walks to the other side of the edge. 
We consider one situation that an \gls{rw} gets bored and 
restarts from a random nodes uniformly(called ``escaping''). 
The escaping probability is denoted by $(1-\alpha)$
and the corresponding stationary distribution
can be solved through the following equation: 
\begin{equation}
	\vec{v} = \alpha \tran{W}\vec{v}
	+ (1-\alpha)\frac{\vec{1}}{||\vec{1}||_1}
	\label{eq:pr1}
\end{equation}
where $\vec{1}$ is the all ones vector, 
and $ 0 < \alpha < 1 $. 

The \gls{pr} algorithm is very successful in 
web ranking context\cite{brin1998anatomy}. 
It is shown to 
satisfy a set of ranking axioms\cite{altman2005pagerank}, 
that fits the web ranking setting very well.
The \gls{rw} interpretation also fits our problem. 
Since our partial topology is biased towards $ o $, 
nodes with higher proximity with $ o $ should 
be visited more frequently by an \gls{rw}. 
If one node has higher proximity with the observer, 
it is more likely to be in the same community. 
This intuition is the same as simple heuristics like
Common Neighbour and Adamic/Adar Score\cite{aggarwal2011social}.  

The initial version of \gls{pr} (\req{\ref{eq:pr0}})
is not informative. 
This is because on an undirected and connected graph, 
the stationary distribution satisfies
$ \vec{v} \propto \vec{d} $. 
However, higher degree does not necessarily mean 
higher proximity with the observer. 
The modified version of \gls{pr} (\req{\ref{eq:pr1}})
re-shares the score among nodes, and results in 
a different ranking. 
On one extreme, when $ \alpha \rightarrow 0 $, 
$ \vec{v} $ is a uniform distribution. 
On the other extreme, when $ \alpha \rightarrow 1 $, 
$ \vec{v} $ is the degree-proportional distribution. 
This version is more informative than the initial version
in \req{\ref{eq:pr0}} 
and we will evaluate how it performs in \rsec{\ref{sec:eval}}.

\subsection{Personalized PageRank}
\glsreset{ev}

Personalized PageRank(PPR), also known as topic sensitive 
pagerank\cite{haveliwala2003topic} is a generalization of 
\req{\ref{eq:pr1}}:
\begin{equation}
	\vec{v} = \alpha W^{\rm T}\vec{v}
	+ (1-\alpha)\frac{{b}}{||\vec{b}||_1}
	\label{eq:ppr}
\end{equation}
where $\vec{b}$ is the personalized \gls{ev}, 
and $ ||\vec{b}||_1 = \sum_i{|\vec{b}_i|} $. 
To simplify the notation, we denote 
$ \vec{\beta} = {b} / {||\vec{b}||_1}$
as the normalized version of $ \vec{b} $. 
Instead of filling all ones in $\vec{b}$, we can make
the escaping probability biased towards a set of nodes. 
The intuition is that the \gls{rw} will restart from 
some nodes with particular interest instead of 
uniformly pick a random node. 
In the stationary distribution, nodes close to 
those restarting nodes will have higher probability. 
If we know some positive nodes beforehand, 
putting them in \gls{ppr}'s \gls{ev} 
will help to rank other positive nodes higher.

\subsection{The Ink Spilling Interpretation of \gls{ppr}}

Besides the random walk view, \gls{ppr} also has a 
nice interpretation using ink spilling process
\cite{berkhin2006bookmark,spielman-2009spectral-ln,lau-2012-spectral-ln}. 
To see this, we reorganize \req{\ref{eq:ppr}} as 
$ (I - \alpha \tran{W})\vec{v} = (1-\alpha)\vec{\beta} $. 
The eigen value of $ \tran{W} = A D^{-1} $ is in the range $ [-1, 1] $, 
so the eigen value of $ (I - \alpha \tran{W}) $ is 
in the range $ [1 - \alpha, 1 + \alpha] $. 
Then $ (I - \alpha \tran{W}) $ is invertible and we have
$ \vec{v} = (I - \alpha \tran{W})^{-1} (1-\alpha)\vec{\beta} $. 
Since the eigen value of $ \alpha \tran{W} $
is in the range $ [ - \alpha, \alpha] $, 
we can perform the expansion $ (I - X)^{-1} = I + X + X^2 + X^3 \ldots $
by taking $ X = \alpha\tran{W}$ in the above equation:
\begin{equation}
	\vec{v} = (1-\alpha) \sum_{t=0}^{\infty}{(\alpha \tran{W})^t} \vec{\beta}.
	\label{eq:ink_spill}
\end{equation}
\req{\ref{eq:ink_spill}} has an intuitive interpretation 
using an ink spilling process:
\begin{enumerate}
	\item Every node $ i $ is initialized
	with $ \vec{\beta}_i $ amount of ink. 
	\item At each step, $ (1-\alpha) $ of the ink 
	dries at every node. 
	\item The rest $ \alpha $ portion wet ink is passed to neighbors 
	using the weighting of $ \tran{W} $. 
	\item The process repeats until all ink dries. 
\end{enumerate}

Based on this ink spilling interpretation, 
we can put some user labeled positive nodes in \gls{ev}. 
Since the positive nodes form a community which 
is well connected inside and have sparse connection 
with the rest network, positive nodes should 
generally have more ink than negative nodes.  
If more positive nodes are put into \gls{ev}, 
\gls{ppr} should be more capable to distinguish
between positive and negative nodes. 
We will numerically investigate the sensitivity of \gls{ppr} to the size 
of restarting set in \rsec{\ref{sec:eval}}.

\subsection{Complexity of \gls{ppr}}
\label{sec:complexity}

Computing \gls{ppr} can be very efficient.
We analyze two methods: Matrix Multiplication
and Ink Spilling.

\subsubsection{Matrix Multiplication}

Based on \req{\ref{eq:ppr}}, we can derive 
the straightforward iteration:
\begin{equation}
	\vec{v}^{({t})} = \alpha \tran{W}\vec{v}^{({t-1})}
	+ (1-\alpha)\vec{\beta} 
	= P \vec{v}^{({t-1})}
	\label{eq:ppr_iter} 
\end{equation}
where $ P = (\alpha \tran{W} + (1-\alpha)\vec{\beta}\tran{\vec{1}}) $. 
The convergence rate of \req{\ref{eq:ppr_iter}} depends 
on the eigen gap of $ P $.
It can be shown that $ \lambda_2(P) \le \alpha $\cite{kamvar2010numerical}, 
which provides a constant gap irrelevant of $ n $. 
Then $ \Theta(\log n) $ iterations are needed to approximate 
the stationary distribution with error bounded 
by $ ||\vec{v}^{({t})} - \vec{v}^{({\infty})}|| = O(\frac{1}{n}) $
\cite{lau-2012-spectral-ln}. 
Basic matrix-vector multiplication costs $ O(n^2) $ time. 
In our implementation, we deal with the two parts of $ P $ separately. 
They cost $ O(m) $ and $ O(n) $ time, respectively. 
The total complexity for a fixed $ \alpha $
is then $ O((m+n)\log n) $
As is shown before, on the 2-hop topology, $ m $ 
is on the order of $ n $, so 
the computation is tractable for every single 
user on commodity desktop computers. 

\subsubsection{Ink Spilling}

Note that the ink spilling process can be performed in an 
asynchronous manner\cite{lau-2012-spectral-ln}. 
We formalize the asynchronous version in \ralg{\ref{alg:ink}}. 
Denote the level of wet ink at node $ i $ by $ \vec{r}(i) $. 
At each step, we pick an arbitrary node 
satisfying the condition $ \vec{r}(i) > \epsilon \vec{d}(i) $, 
dry $ (1-\alpha) $ portion of wet ink, and 
share the remaining  ink among its direct neighbors. 
If no such node exists, i.e. $ \vec{r}(i) \le \epsilon \vec{d}(i), \forall i $, 
we call the current $ \vec{v_{\epsilon}} $ an $ \epsilon $-approximation 
of \gls{ppr}. 
The termination condition implies that there is only a 
small fraction of wet ink at all nodes. 

\begin{algorithm}[!t]
	\caption{Ink Spilling Algorithm}
	\label{alg:ink}
	\begin{algorithmic}[1]
		\REQUIRE Walk matrix: $ W $
		\REQUIRE Escape vector: $ \vec{b} $
		\ENSURE $ \epsilon $-approximation: $ \vec{v_{\epsilon}} $
		\STATE $ \vec{r} \leftarrow \vec{\beta}  $
		\WHILE{$ \exists i, \text{s.t.} \vec{r}(i) > \epsilon \vec{d}(i) $} 
		\STATE $ \vec{v_{\epsilon}}(i) 
		\leftarrow \vec{v_{\epsilon}}(i) + (1-\alpha) \vec{r}(i)$
		\STATE $ \forall j \in N^{(1)}_i,  \vec{r}(j) \leftarrow
		\vec{r}(j) + \alpha W_{ij} \vec{r}(i)  $
		\ENDWHILE
	\end{algorithmic}
\end{algorithm}

We denote the node selected at step $ t $ by $ v_t $. 
From the iteration invariance, we know
at least $ \epsilon \vec{d}(v_t) $ ink is wet at this node. 
Since the initial ink level is $ \sum_i\vec{\beta}_i = 1$,
we have $ \sum_t{(1-\alpha) \epsilon \vec{d}(v_t)} \le 1 $. 
That is $ \sum_t{ \vec{d}(v_t)} \le 1/((1-\alpha) \epsilon) $. 
In each step, it takes constant time to dry a portion of ink for the current node
and $ \vec{d}(v_t) $ time to distribute wet ink to neighbors. 
So the complexity of ink-spilling algorithm is 
$ \sum_t{ \vec{d}(v_t)} = O(\frac{1}{(1-\alpha) \epsilon}) $ \cite{lau-2012-spectral-ln}. 
To reach the same level of approximation as 
matrix multiplication, 
i.e. $ \sum_i \epsilon \vec{d}(i) = \Theta(\frac{1}{n}) $, 
we have $ \epsilon = \Theta(\frac{1}{mn}) $. 
As $ \alpha $ is constant for a given graph size, 
the overall complexity is $ O(mn) $. 
Note that in the real application, performing excellent 
approximation is not necessary. 
So we can set $ \epsilon $ to a larger value than $ \Theta(\frac{1}{mn}) $. 
We also note that this analysis is not tight
and in practice ink spilling algorithm can benefit 
from the sparse structures of many graphs. 

\section{Performance Evaluation}
\label{sec:eval}

 We have evaluated our proposed algorithms using 
data sets crawled from Renren,
currently the largest \gls{osn} in China.

\subsection{Data Set}
\label{sec:data}

By default, a Renren user can only view the buddy list 
of their direct friends. This setting incidentally restricts the network perspective of an individual
Renren user to his/her 2-hop neighborhood, matching exactly the scenario we considered in 
\rsec{\ref{sec:form}}. We have developed a dedicated crawler to collect 
2 hop buddy list information
(friends and friends of friends)
from any given observer. 
Eight volunteers have helped 
to run the crawler and contributed their own data. 
All the volunteers are active 
undergraduate Renren users for classes 2006 to 2008. 
Our study is based on the separately anonymized 
version of their data. 
As our formulation only relies on 2-hop topology 
of any target user,  we actually have collected eight independent data sets
for evaluation.  
Since the results for these eight data sets are qualitatively the same, in what follows,
we only present the quantitative evaluation results for  one observer's data set
(denoted by $ G^{(2)}_o $) for brevity. 

Manual examination of the data sets has reviewed that one's largest and also latest community 
always (in our data sets) correspond to his/her undergraduate university's community. 
In this initial evaluation, we are interested in the ability of our PPR approach in ``re-discovering''
this community based on 2-hop topology. 
For observers who have changed their institution name 
after graduation, we manually set 
his/her institution name to the undergraduate university. 
For any other nodes, we stick to the 
default institution name shown on the buddy list.
All nodes that have the same institution name as the observer 
are regarded as positive nodes, and the rest are regarded as 
negative nodes. 
This preprocessing causes certain noise of the ground-truth 
but this is the best we can do
for it's impossible for us to get all institution names of a person by default.

\subsection{Evaluation Methodology}

\glsreset{roc}
\glsreset{auc}

In \rsec{\ref{sec:form}}, we listed some conventional 
evaluation criterion for classification problem. 
However, simple evaluation like accuracy can be misleading. 
For instance, in our experiment,  observer $ o $ has
88238 nodes in his 2-hop neighborhood (L1 + L2 friends). 
9345 of them are positive nodes
and the ratio of is 0.1059. 
That means, the most naive algorithm which declares all 
nodes to be negative can reach an 
accuracy(\req{\ref{eq:def_accuracy}}) of 0.8941. 
This result is not of practical interest. 
Instead of accuracy, 
we look at \gls{tpr} and \gls{fpr} separately. 

In our two-stage algorithm framework, 
we first calculate a ranking of all nodes in $ T_o $
and then set a proper threshold $ R $ to cut the set 
into positive and negative set. 
By varying the threshold, we can obtain 
a series of \gls{tpr} and \gls{fpr}. 
Plotting all those points on a 2-D graph, 
we can get the \gls{roc} curve\cite{fawcett2006introduction}. 
If we rank nodes randomly, 
the ROC curve is approximately a straight line connecting (0,0) and (1,1). 
The higher the \gls{roc} curve is above the diagonal line, 
the more capable a ranking is to distinguish
between positive and negative nodes. 

Although \gls{roc} is very informative for human,
it is hard to automatically compare many \gls{roc} curves directly.  
Instead, the \gls{auc}\cite{fawcett2006introduction} is
a better summary metric. 
If the \gls{auc} is larger, the \gls{roc} curve is generally higher. 
\gls{auc} is equivalent to 
the probability that an algorithm will rank 
a randomly chosen positive node 
higher than a randomly chosen negative node\cite{fawcett2006introduction}.

\subsection{Validation of \gls{auc} as An Effective Metric}

Although \gls{auc} reflects the expected 
performance of a classifier, 
it does not mean higher \gls{auc} curves will 
dominate lower \gls{auc} curves everywhere. 
In this section, we experimentally show that 
\gls{auc} is an effective metric. 

\begin{figure}[!t]
	\centering
	\includegraphics[width=0.6\linewidth]{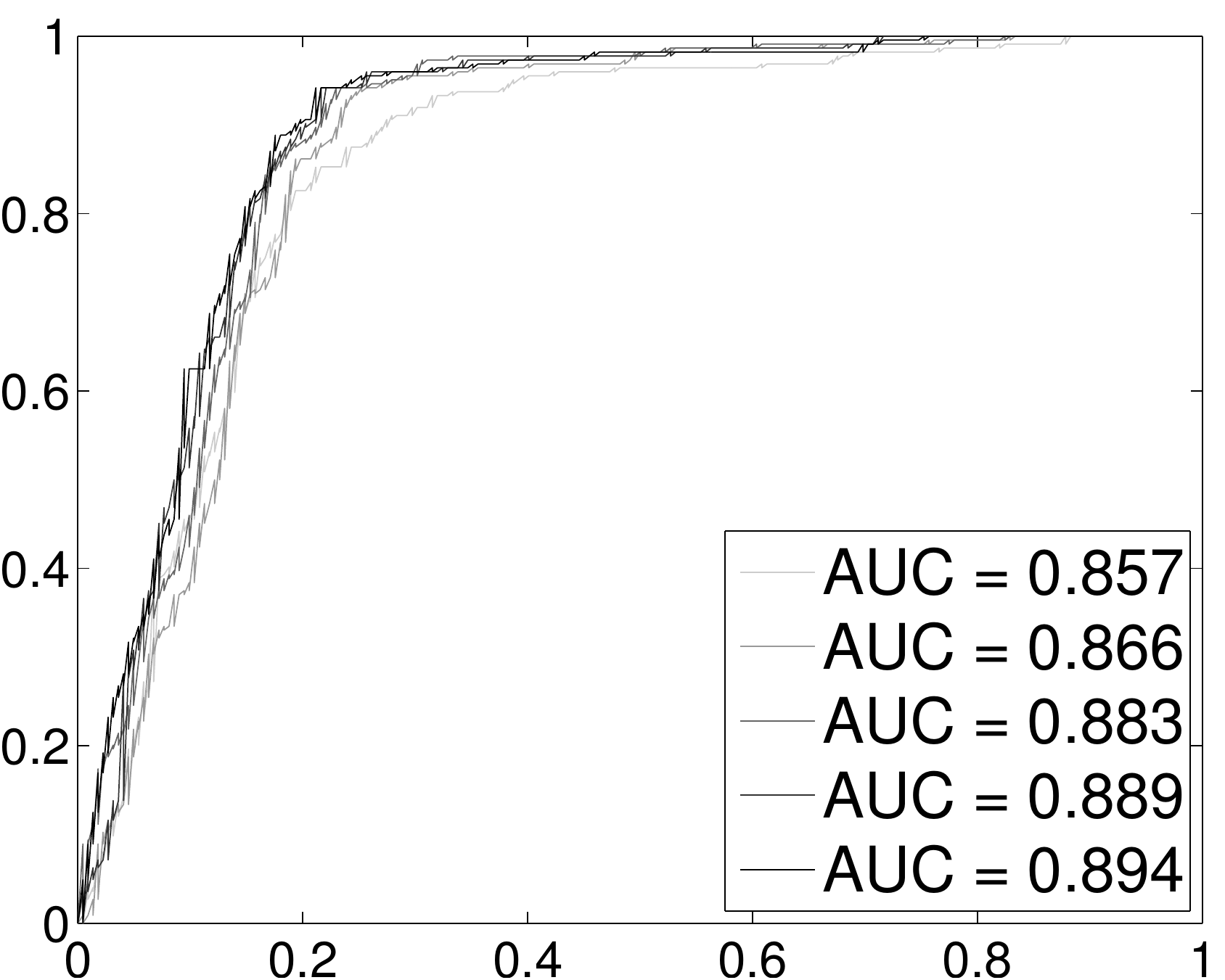}
	\caption{\gls{roc} Curves with \gls{auc} Increases from Light to Dark}
	\label{fig:case_rnd_l1}
\end{figure}

In this experiment, we configure \gls{ev} 
of \gls{ppr} randomly. 
We run 10 instances of simulations and 
plot the \gls{roc} curves in \rfig{\ref{fig:case_rnd_l1}}. 
The test set $ T_o $ is set to L1 nodes ($ T_o = N^{(1)}_o $). 
To make the plot uncluttered, 
only 5 of the curves are plotted. 
It is shown that larger \gls{auc} in general 
implies higher \gls{roc} curve. 
This is partly because those curves
come from the same family, i.e. \gls{ppr} value
under different settings.
 
For the rest of our evaluation, 
we will focus on comparing the \gls{auc}
of different approaches. 

\subsection{Leverage \gls{roc} to Get Classification Result}

Getting a good \gls{roc} curve is just an intermediate step 
towards the final application. 
Before we extensively evaluate \gls{auc} of different 
approaches under different settings, 
we briefly note two methods to leverage the 
\gls{roc} curve to get classification result. 

Suppose we get one \gls{roc} curve in \rfig{\ref{fig:case_rnd_l1}}. 
One attainable ROC point (0.1832,0.8991) 
can be read from the plot. 
In $ N^{(1)}_o $, there are 224 positive nodes 
and 222 negative nodes. 
This means that 
an algorithm can return $ 0.1832 \times 222 = 40.670 $ negative nodes
and $ 0.8991 \times 224 $ = 201.40 positive nodes with proper threshold.
The problem now becomes how to target 
this \gls{roc} point without knowing the ground-truth. 
Here are two choices:
\begin{itemize}
	\item With prior knowledge of the portion of positive nodes
	(obtained through statistics over the network), 
	we can target an \gls{roc} point in the following way. 
	For example, suppose we find 
	(0.19,0.9) is a reasonable \gls{roc} point to target. 
	For a typical undergraduate user, he/she usually 
	have half of the friends come from the same university. 
	Then we vary threshold $ R $ until the number of predicted 
	positive nodes reaches 
	$ 0.19 \times 0.5 \times |N^{(1)}_o| + 0.9 \times 0.5 \times |N^{(1)}_o|
	= 0.545 |N^{(1)}_o| $. 
	\item Another method is to look at \gls{ppr} values 
	directly and detect possibly existing sharp drops. 
	Andersen\cite{andersen2007detecting} studied the 
	local graph partitioning problem and proposed to 
	cut the graph by detecting sharp drops of the \gls{ppr} vector. 
	We already note the difference between Local Graph Partitioning 
	and the ``localized'' setting of our problem 
	in \rsec{\ref{sec:intro}}. 
	Whether similar approaches are applicable to the current 
	problem is left to future work. 
\end{itemize}

\subsection{Comparison of Four Approaches}
\label{sec:roc_4f}

We first compare our 
\gls{pr} and \gls{ppr} proposal
to two local heuristics:
\begin{itemize}
	\item Common Neighbour (\req{\ref{eq:common}}) simply 
	computes the size of intersection between one node's
	neighborhood and the observer's neighborhood. 
	The intuition is that the more friends the two nodes have 
	in common, the more likely they are in the same community. 
		\begin{equation}
			\text{Common}(i,j) = | N^{(1)}_i \cap N^{(1)}_j |
			\label{eq:common}
		\end{equation}
	\item Adamic/Adar score (\req{\ref{eq:adar}})
is an intuitive improvement of Common Neighbour. 
It biases towards lower-degree nodes. 
The reason is that higher degree nodes are more popular
and thus provide little information if two nodes 
both connect to it. 
On the contrary, if two nodes both connect to a low-degree 
node, they have a larger chance of being in the same community. 
		\begin{equation}
			\text{Adamic/Adar}(i,j) = \sum_{k \in N^{(1)}_i \cap N^{(1)}_j }
			{\frac{1}{\log{|N^{(1)}_k|}}}
			\label{eq:adar}
		\end{equation}

\end{itemize} 

$ f(i) = \text{Common}(o,i) $ and $ f(i) = \text{Adamic/Adar}(o,i) $
are computed to rank node $ i $, respectively. 
The two simple heuristics  
can be computed exactly within 2-hop topology
and are thus the right baseline to compare.
For many other algorithms, their original settings
are beyond 2-hop. 
Running those algorithms on our data set naturally violates 
their assumptions so the performance is not guaranteed. 

In this experiment, the \gls{ev} $\vec{b}$ of \gls{ppr}
is simply set to 3 highest degree positive nodes plus
the observer $ o $. 
We defer the exploration of how the choice of $\vec{b}$ 
would influence the performance to following sections. 
For the basic version of \gls{pr}, 
the \gls{ev} is set to  an all-one's vector. 
Both \gls{ppr} and \gls{pr} use a transition ratio of $ \alpha=0.9 $. 
The results of the four approaches are evaluated 
on all nodes within 2-hop, i.e. $ T_o = N^{(2)}_o $.

\begin{table}[!t]
	\centering
	\caption{\gls{auc} of Four Different Approaches}
	\label{tbl:auc_4f}
	\begin{tabular}{|c|c|c|c|}
		\hline
		Common & 0.7415 & Adamic/Adar & 0.7428 \\
		\hline
		PR & 0.7024 & PPR & 0.8339 \\
		\hline
	\end{tabular}
\end{table}

\begin{figure}[!t]
\centering
\includegraphics[width=0.7\linewidth]{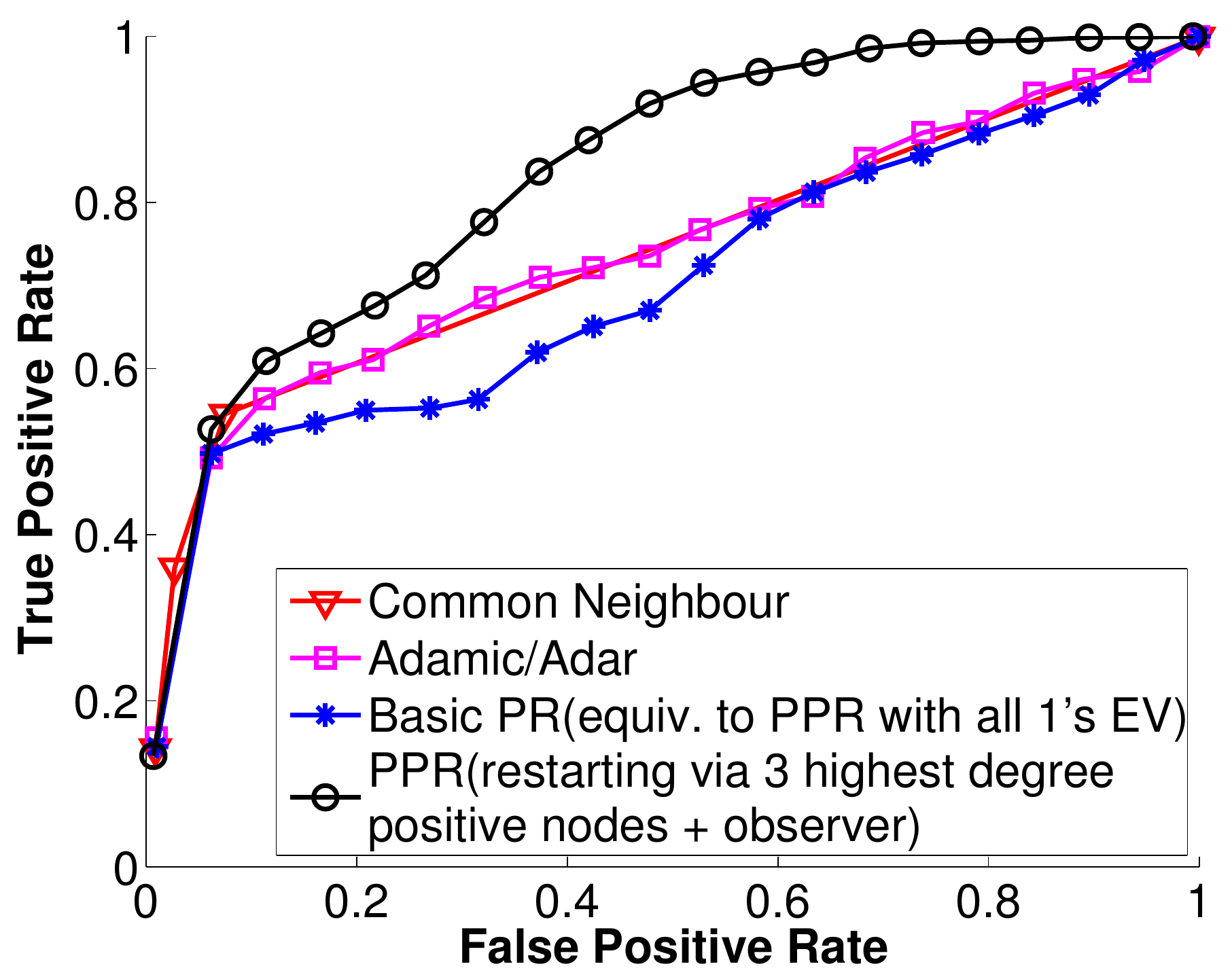}
\caption{ROC Curve of Four Different Approaches}
\label{fig:roc4f}
\end{figure}

\rtbl{\ref{tbl:auc_4f}} shows the \gls{auc} of 
the four approaches and
\rfig{\ref{fig:roc4f}} shows their \gls{roc} curves.
We can see that Common Neighbours 
and Adamic/Adar perform equally well. 
In the low \gls{fpr} region
(e.g. $ \text{FPR} < 0.1 $ ), the four approaches 
does not have much difference. 
In the higher \gls{fpr} region, 
\gls{ppr}
is obviously more effective 
than the others while pure \gls{pr}
even does worse than the two simple heuristics. 
Note that the \gls{auc} is at least 0.5 for meaningful 
approaches, \gls{ppr} actually improves \gls{pr} by 64.97\% relatively. 
This result aligns with our analysis in \rsec{\ref{sec:algorithm}}.

\begin{myre}
\textbf{A few labels help a lot}. 
	If the user can manually label a few positive nodes, 
	the \gls{ppr} algorithm can leverage 
	this information to yield significant improvement. 
\end{myre}

\subsection{Evaluation on All Nodes within 2-Hop}

We already observed that a slight modification 
of \gls{ev} $ \vec{b} $ can make a considerable difference 
in the \gls{roc} curve, 
whereas simple-minded all one's $ \vec{b} $
performs worse than the two local heuristics. 
In the previous proof-of-concept experiment, 
we put 3 highest degree
positive nodes and the observer in the \gls{ev}. 
In this section, we explore how \gls{auc} 
varies with different choices of $ \vec{b} $. 

Among different alternatives to fill in $ \vec{b} $, 
we choose the unweighted version of $ \vec{b} $ for simplicity. 
That is, we determine a restarting set of nodes, denoted by 
$ S \subset N^{(2)}_o $, 
and fill in $ \vec{b} $ as follows:
\begin{equation}
	\vec{b}_i = \left\lbrace
	\begin{array}{cc}
	1 & i \in S \\
	0 & i \notin S
	\end{array}
	\right.
\end{equation}

\subsubsection{Random Positive Nodes as Restarting Set}

\begin{figure}[!t]
	\centering
	\subfigure[AUC v.s. $ |S| $,50 Rounds]{
		\includegraphics[width=0.45\linewidth]{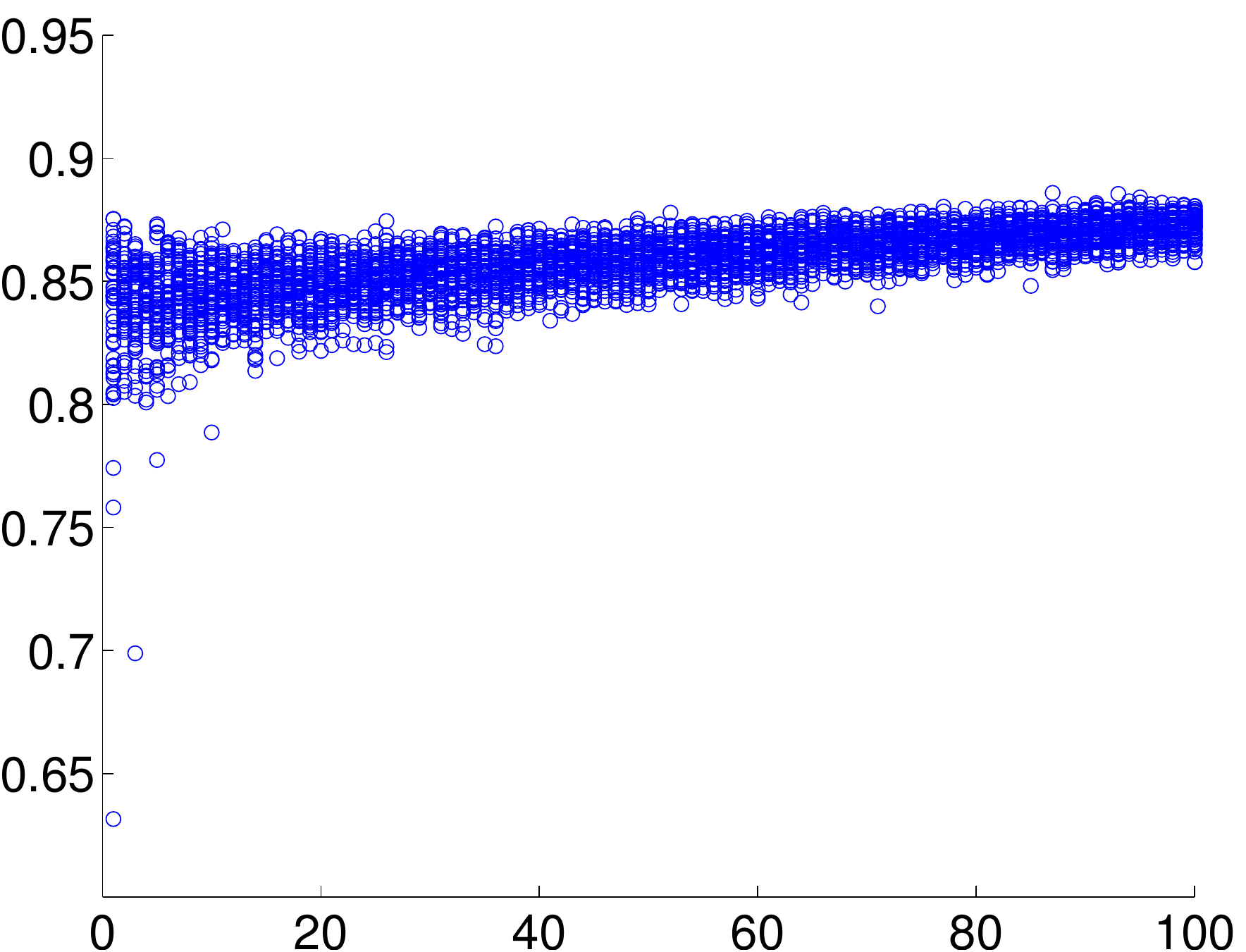}
		\label{fig:auc_rnd_full_all}
	}
	\subfigure[Mean of AUC v.s. $ |S| $]{
		\includegraphics[width=0.45\linewidth]{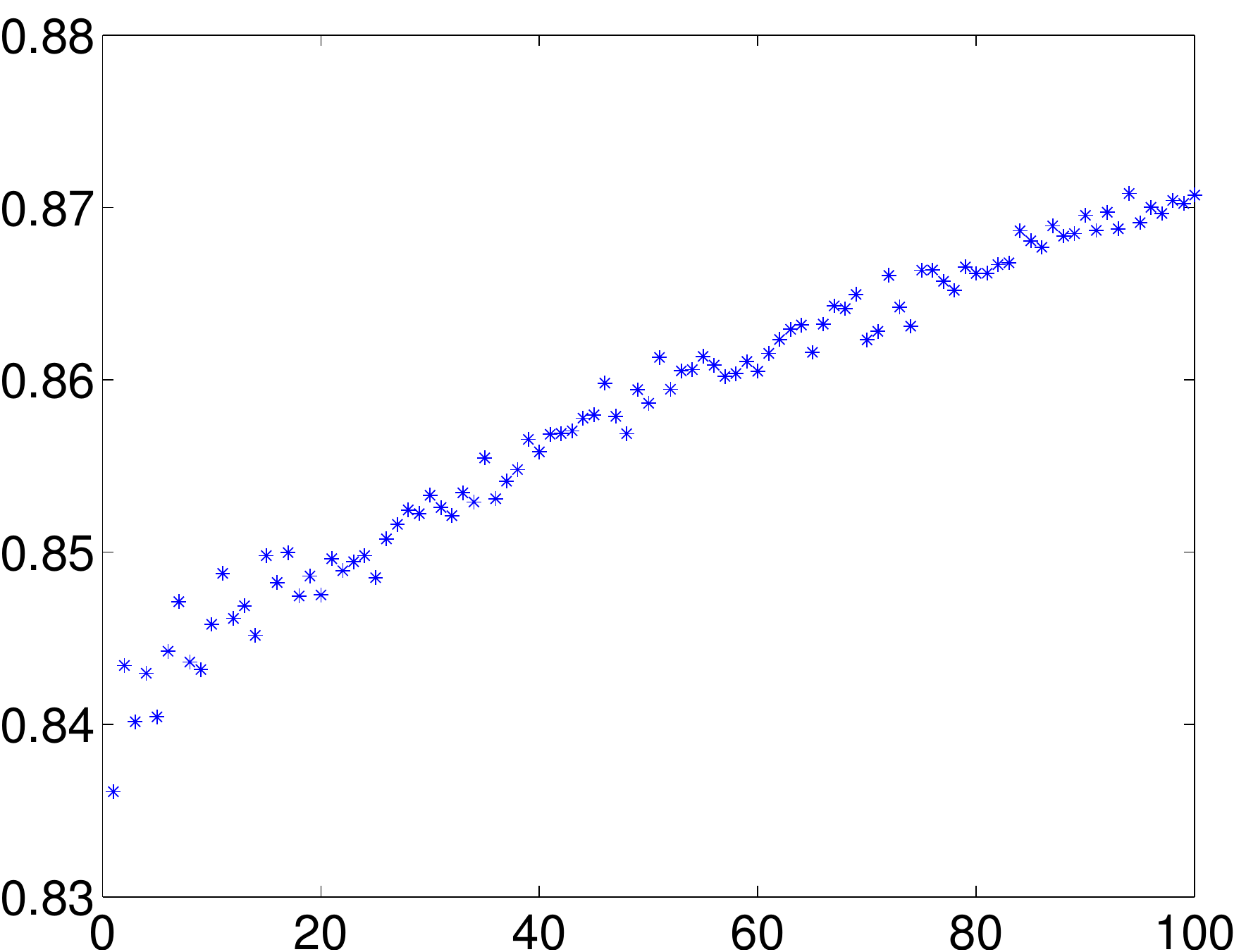}
		\label{fig:auc_rnd_full_mean}
	}
	\caption{Random $ S $, \gls{auc} Evaluated on $ N^{(2)}_o $}
	\label{fig:auc_rnd_full}
\end{figure}

In this experiment, some random positive nodes are put into $ S $. 
For each size of $ S $, we repeat the experiment for 50 rounds
to absorb randomness. 
\rfig{\ref{fig:auc_rnd_full_all}} is the scatter 
plot of 5000 simulations with $ |S| $ varying 
from 1 to 100. 
\rfig{\ref{fig:auc_rnd_full_mean}} plots the mean 
\gls{auc} of each $ |S| $.
It is clear that more pre-known positive nodes 
help to improve the performance of \gls{ppr} 
on our 2-hop community detection settings
but there is only a little benefit
($ 0.03 $ increase of \gls{auc} from $ |S|=1 $ to $ |S|=100 $, i.e. an
8.8\% relative improvement). 
Besides the increase of mean, larger 
$ |S| $ results in smaller variance, which 
is also appreciated in real applications.
Manual labeling causes a cost for the users,   
so they need to find a trade-off between performance and cost. 

\subsubsection{High-Degree Positive Nodes as Restarting Set}

\begin{figure}[!t]
	\centering
	\subfigure[Evaluated on $ N^{(2)}_o $]{
		\includegraphics[width=0.45\linewidth]{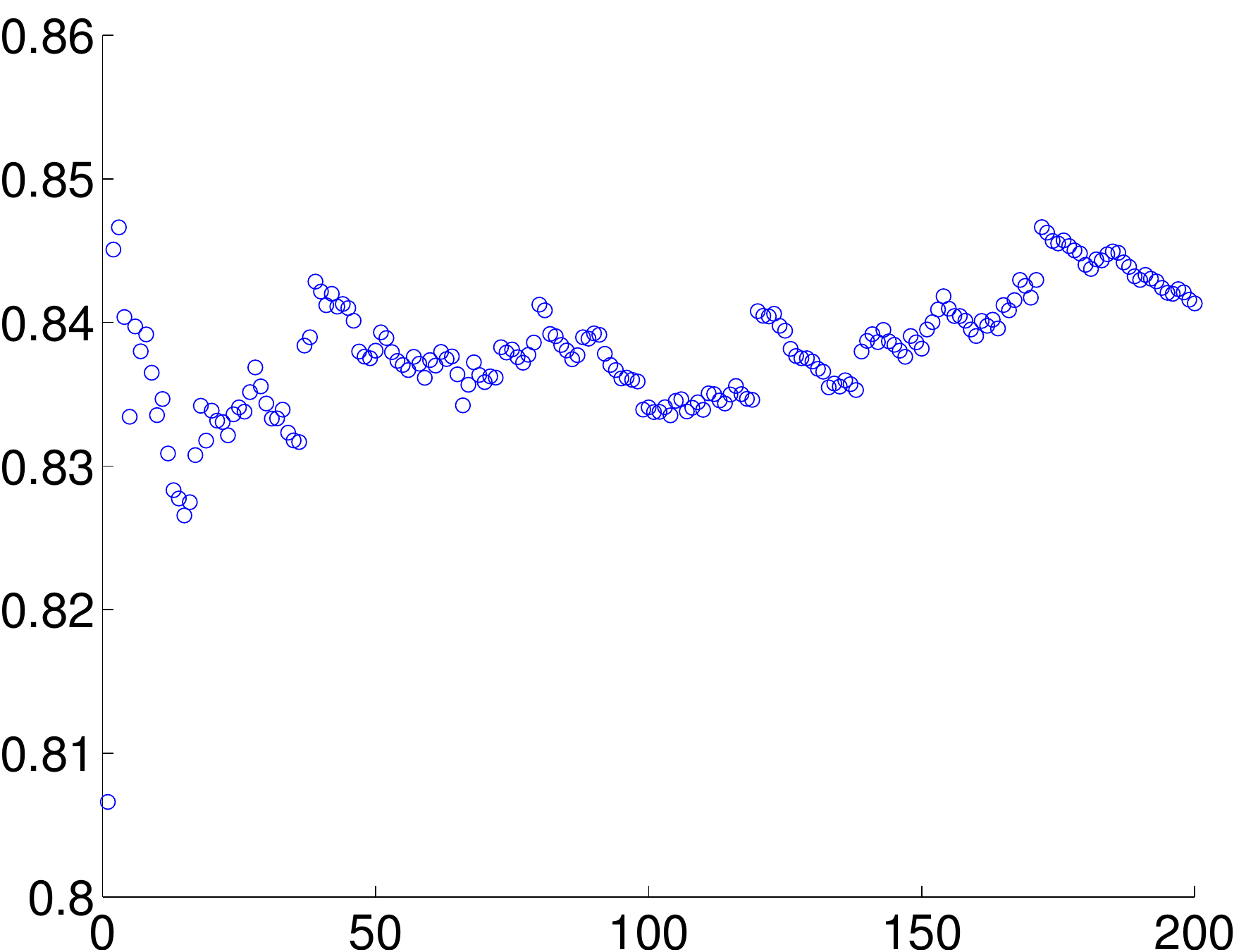}
		\label{fig:auc_highd_all}
	}
	\subfigure[Evaluated on $ N^{(1)}_o $]{
		\includegraphics[width=0.45\linewidth]{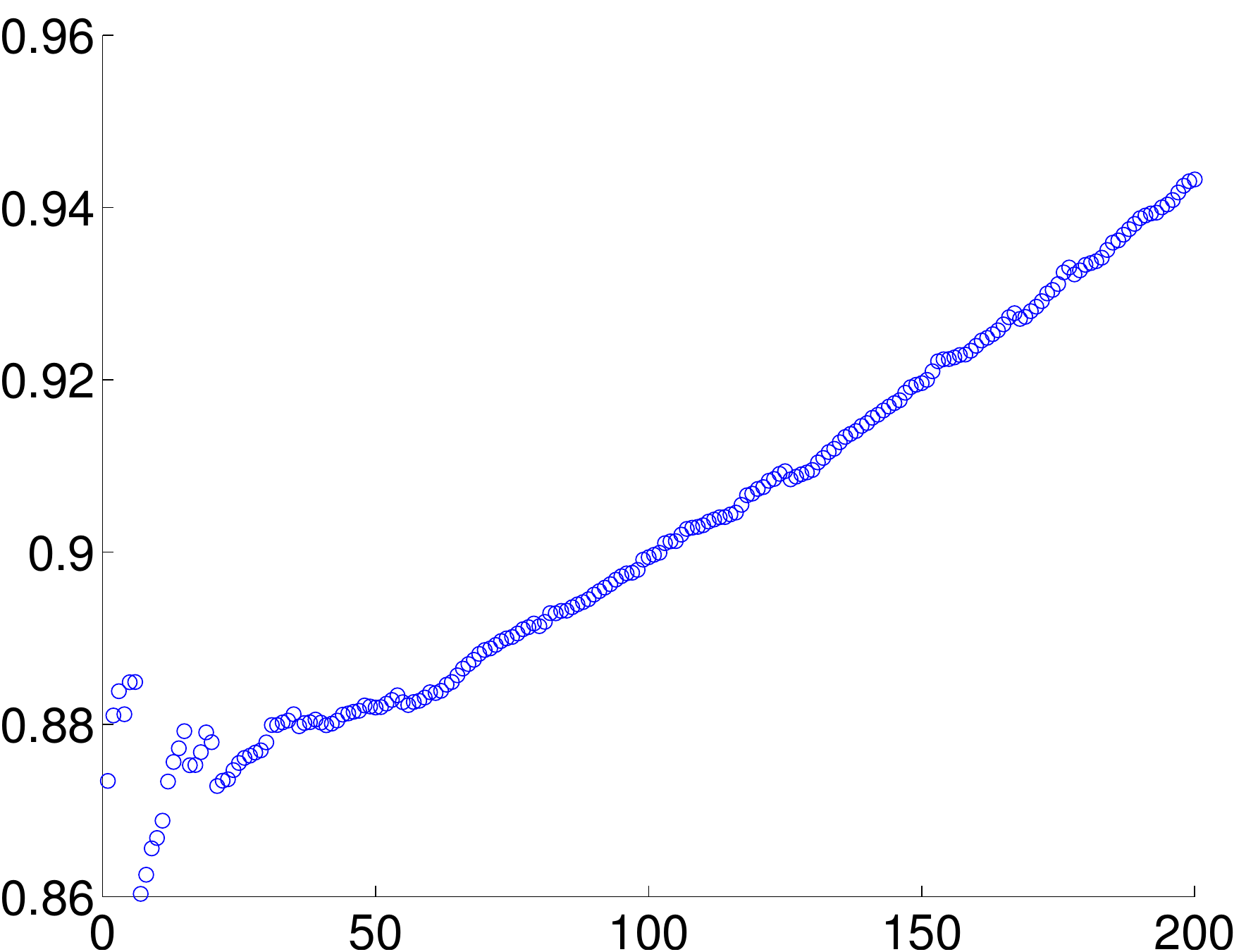}
		\label{fig:auc_highd_level1}
	}
	\caption{\gls{auc} v.s. Number of High Degree Nodes}
	\label{fig:auc_highd}
\end{figure}

In this experiment, 
the $ x $ highest degree positive nodes are put into $ S $. 
We vary $ x $ from 1 to 200. 
\rfig{\ref{fig:auc_highd_all}} shows the scatter plot 
of \gls{auc} v.s. $ x $. 
There is no obvious pattern in this plot and 
high-degree heuristic generally performs
worse than random positive node heuristic
when $ |S| $ is fixed. 
As a practical note, for the application on $ N^{(2)}_o $
we suggest the user randomly label a few  positive nodes.

\begin{myre}
\textbf{More labels help a little}. 
	In the random setting, 
	putting more nodes in the \gls{ev} results
	in minor linear benefit. 
	In the high degree setting, 
	more nodes does not improve the performance.
\end{myre}

\subsection{Evaluation on Only Level-1 Nodes}

In this section, we repeat the same experiment 
in the previous section but evaluate the result 
on $ N^{(1)}_o $, i.e. only L1 friends. 
The detection result on $ N^{(1)}_o $ cannot 
be used to do friend recommendation, 
for those nodes are already connected to observer. 
Nevertheless, it is still meaningful 
for automated relationship categorization. 

\subsubsection{Random Positive Nodes as Restarting Set}

\begin{figure}[!t]
	\centering
	\subfigure[AUC v.s. $ |S| $,50 Rounds]{
		\includegraphics[width=0.45\linewidth]{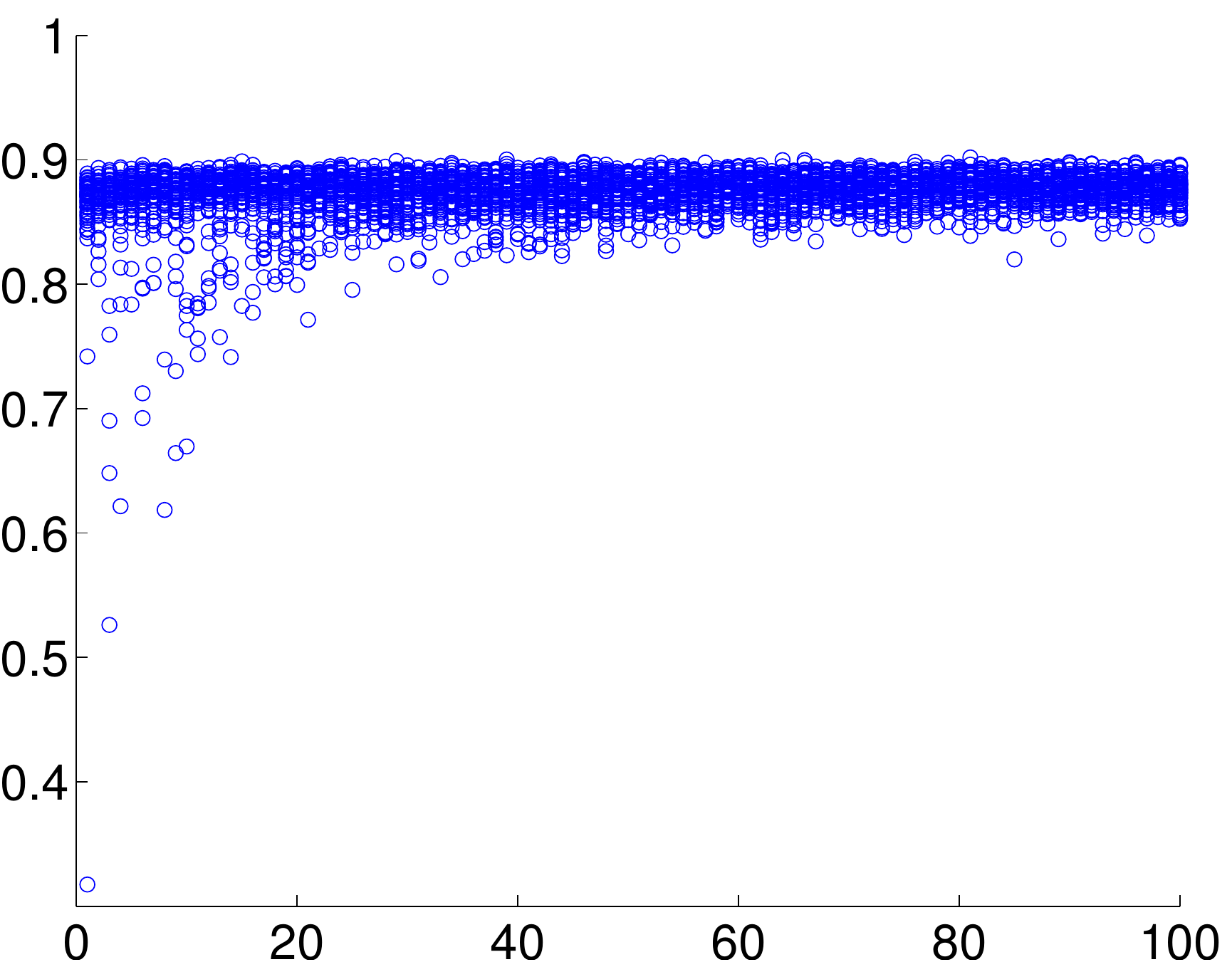}
		\label{fig:auc_rnd_level1_all}
	}
	\subfigure[Mean of AUC v.s. $ |S| $]{
		\includegraphics[width=0.45\linewidth]{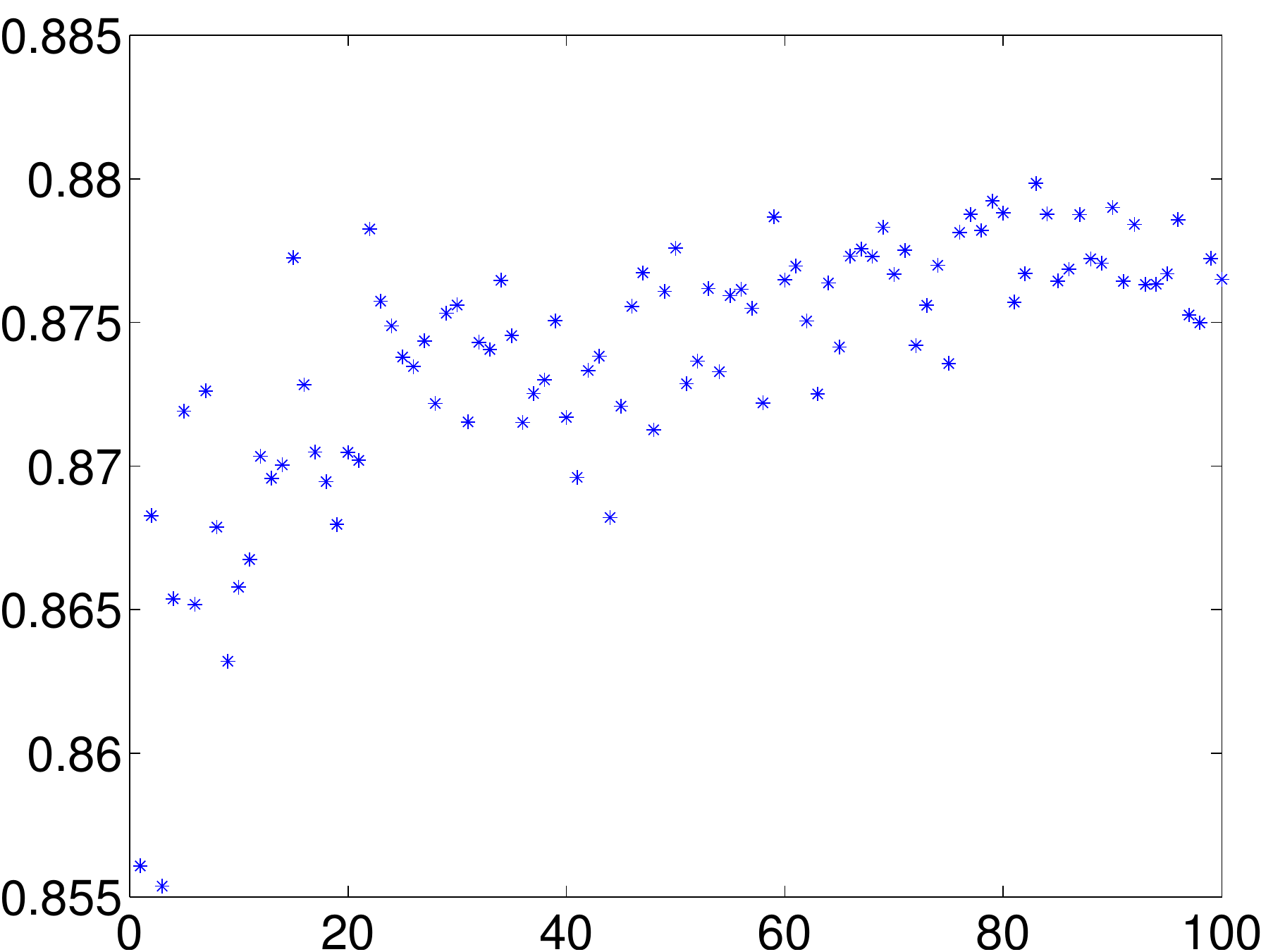}
		\label{fig:auc_rnd_level1_mean}
	}
	\caption{Random $ S $, \gls{auc} Evaluated on $ N^{(1)}_o $}
	\label{fig:auc_rnd_level1}
\end{figure}

We first choose the random $ S $, and plot 
two similar graphs as before in 
\rfig{\ref{fig:auc_rnd_level1}}. 
It shows that the mean of \gls{auc} does
not increase with $ |S| $ 
but the variance decreases with $ |S| $. 
Differences in \rfig{\ref{fig:auc_rnd_level1_mean}}
are mainly due to randomness. 

\subsubsection{High-Degree Positive Nodes as Restarting Set}

Next we evaluate high-degree heuristic
and plot the result in \rfig{\ref{fig:auc_highd_level1}}. 
After a fluctuation period, the \gls{auc}
goes up steadily with the increase of $ |S| $. 
Note that we are evaluating \gls{ppr} on L1 nodes 
in this experiment. 
On the 2-hop topology seen by our observer, 
the high degree positive nodes are also mostly L1 nodes. 
Consider the ink spilling process. 
More L1 positive nodes in $ S $ means they have more initial ink. 
At least, nodes in $ S $ are guaranteed to have higher rank. 
The result of \rfig{\ref{fig:auc_highd_level1}} 
is thus intuitive. 
Note that the curve in \rfig{\ref{fig:auc_highd_level1}}
waits until $ |S| \approx 60$ to reach the linear increasing part. 
Labeling 60 highest degree positive nodes 
is a prohibitive workload for a normal user
and is thus unrealistic in practice.
 
Comparing \rfig{\ref{fig:auc_highd_level1}} 
with \rfig{\ref{fig:auc_rnd_level1}}, 
the high degree positive heuristic generally performs better 
than random node heuristic. 
As a practical note, for the application on $ N^{(1)}_o $, 
we suggest labeling several high degree positive nodes. 
High degree nodes in a community are not only easy to recognize, 
but also induce a deterministic construction of \gls{ev}.  

\begin{myre}
\textbf{$ G^{(h)}_o $ is the minimum for
	the application on $ N^{(h-1)}_o $}.
	Although \gls{ppr} shows significant improvement to simple heuristics, 
	the result on all 2-hop nodes is not good enough for practical use. 
	This is shown by the ROC curves in 
	\rfig{\ref{fig:case_rnd_l1}} and \rfig{\ref{fig:roc4f}}.
	The evaluation on only level 1 nodes is promising, 
	and we can find good  \gls{roc} point to target 
	in real applications. 
\end{myre}

\subsection{Convergence Behavior of Two \gls{ppr} Algorithms}
\label{sec:runtime}

In this section, we evaluate the runtime behavior
of the two candidate realizations of \gls{ppr}:
Matrix Multiplication and Ink Spilling. 
The data set is from the same observer as above sections. 
To show the convergence behavior, we fix 
$ \alpha = 0.9 $ and 
set the \gls{ev} using the 5 highest degree positive nodes. 
By varying $ \epsilon $ for the two algorithms, 
we obtain the convergence rate and error. 
All computations are done on a laptop 
with 2.00 GHz CPU. 

\begin{figure}[!t]
	\centering
        \subfigure[Matrix Multiplication]{%
            \label{fig:com_matrix}
            \includegraphics[width=0.49\linewidth]{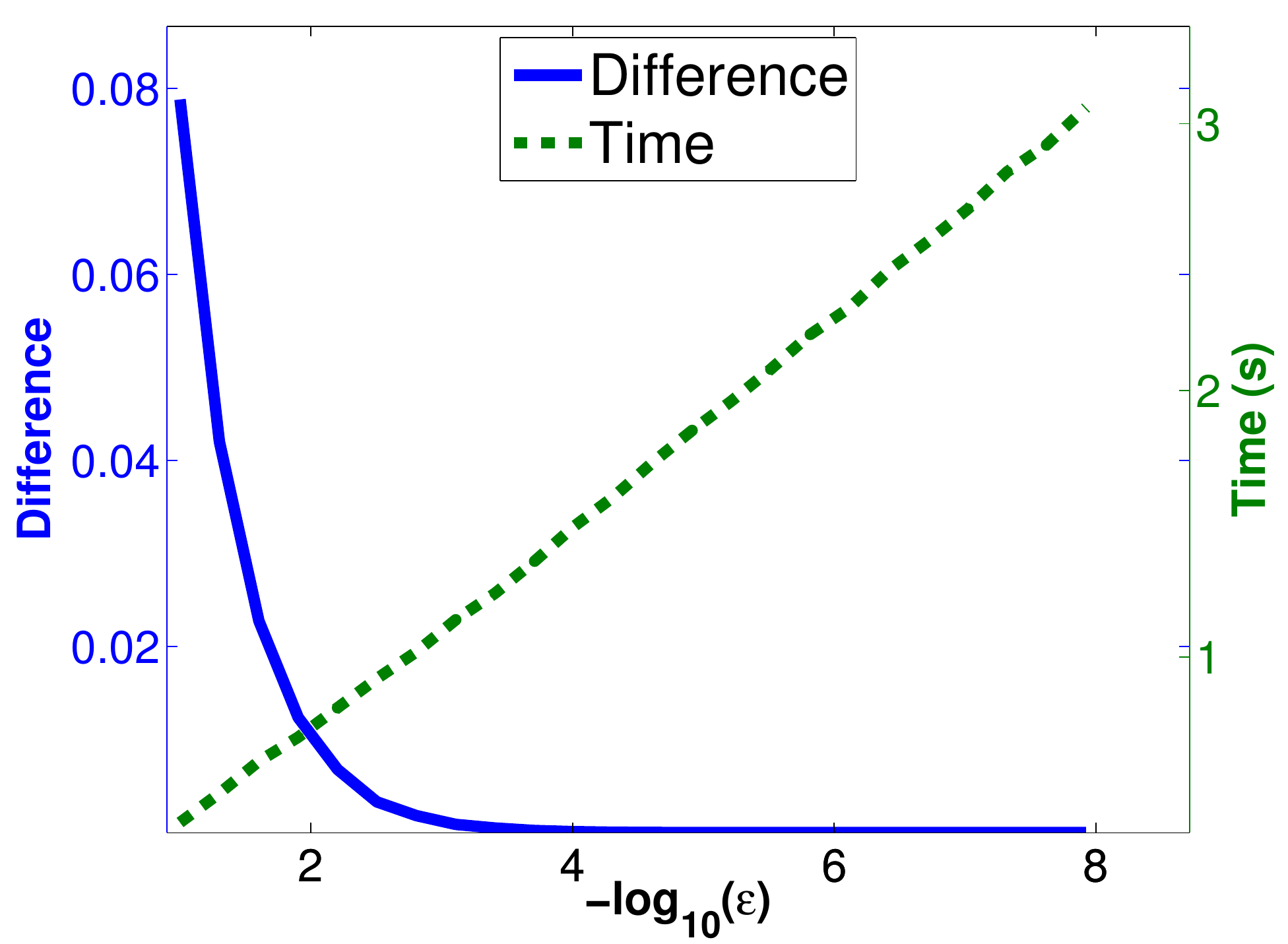}
        }%
        \subfigure[Ink Spilling]{%
            \label{fig:com_ink}
            \includegraphics[width=0.49\linewidth]{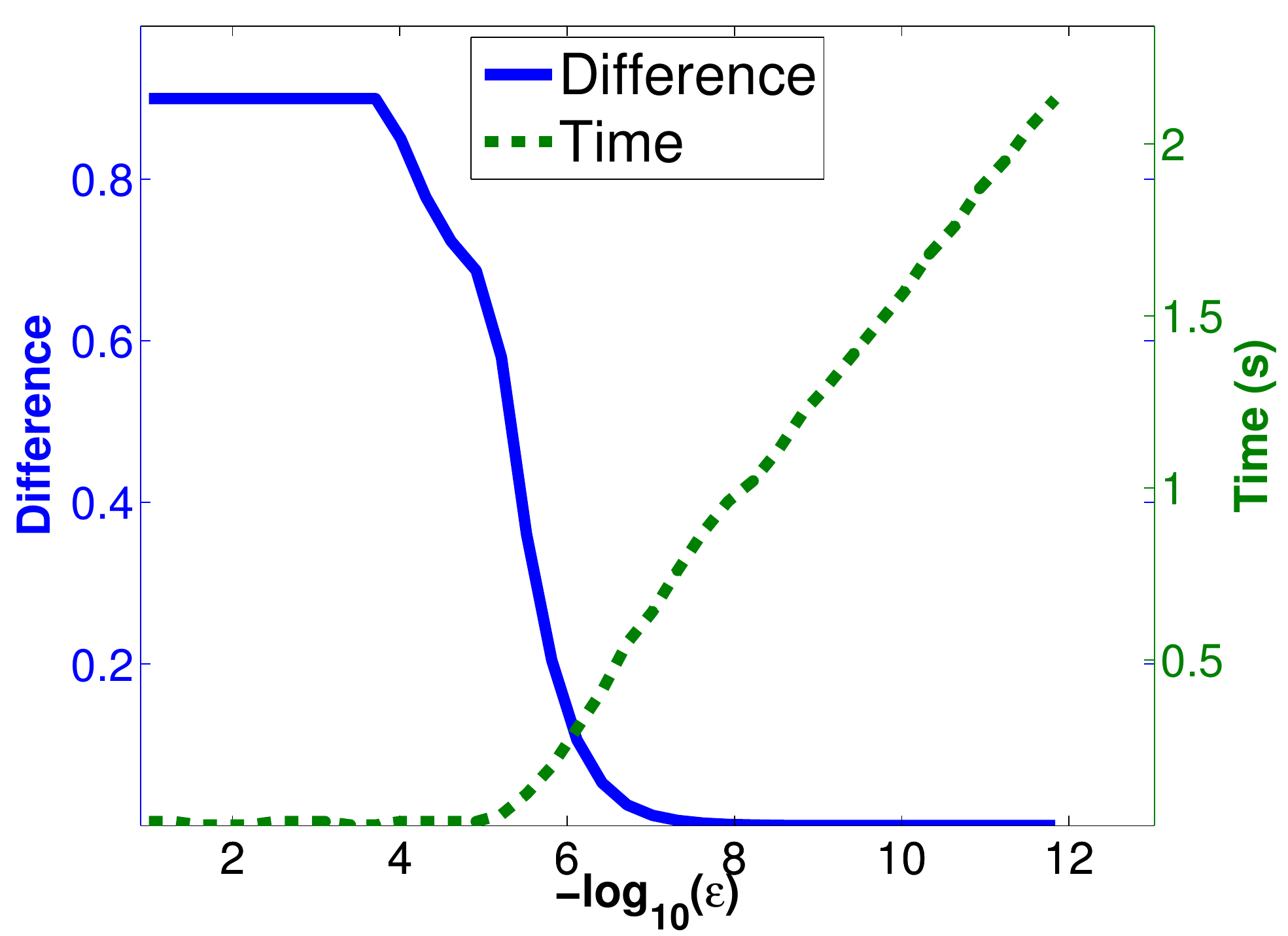}
        }%
    \caption{Convergence Behaviour of Two Algorithms}
    \label{fig:com}
\end{figure}

Note that the analysis of matrix multiplication algorithm 
in \rsec{\ref{sec:complexity}} does not depend on 
the choice of vector norm. 
We use $ ||.||_1 $ in our implementation
for both simplicity and numerical stability.
Denote the result of algorithm ``algo'' with 
precision control parameter $ \epsilon $ by 
$ \vec{v}^{\text{(algo)}}|_{\epsilon} $. 
We choose $ \vec{v}^{\text{(bm)}} = \vec{v}^{\text{(matrix)}}|_{\epsilon=10^{-10}} $
as the benchmark. 
We record the difference 
$ ||\vec{v}|_\epsilon -  \vec{v}^{\text{(bm)}}||_1$
and execution time for each $ \epsilon $. 
\rfig{\ref{fig:com}}
plots the convergence behaviour of 
the two algorithms.

The x axis($ -\log_{10}\epsilon $) in \rfig{\ref{fig:com}} is in log-scale. 
We observe a linear part in both plots.
This means the two algorithms both converges exponentially fast. 
In around 2 seconds, both of the algorithms can converge 
to a ranking function with 
$ ||\vec{v}|_\epsilon -  \vec{v}^{\text{(bm)}}||_1 < 10^{-6} $, 
which is more than enough for  most practical purposes. 
The flat part of the curve in 
\rfig{\ref{fig:com_ink}} also indicates that
more stringent $ \epsilon $ should be set 
in order for the Ink Spilling approach to converge. 
This is because the definition of $ \epsilon $-approximation 
in the Ink Spilling algorithm involves node degree. 
For example, suppose we have
a local topology with $ |E(N^{(2)}_o)| \approx 10^{5} $, 
at least $ \epsilon \approx 10^6 $ is required to make the total 
error upper bounded by $ 0.1 $ when the algorithm converges. 

\section{Discussions and Future Works}
\label{sec:discussion}

The extensive simulations above show that PPR is an 
effective and flexible proximity measure. 
With appropriate parameters, the PPR score can be used 
as a reasonable ranking function for our binary classification framework. 
We have also remarked that the application of PPR on L1 nodes are more appropriate for real application. 
There are good ROC points to target and we mentioned some methods,
e.g. using prior distribution or detecting sharp drop
,to target them. 
We illustrated the use of prior distribution. 
However, it is not trivial to set proper threshold in order to get final detection result. 
We leave this as a future work. 

We have also remarked that 2-hop topology is not enough for detecting community of all 2-hop nodes (L1 + L2).
Meanwhile, 2-hop topology is promising for community detection of L1 nodes. 
If we constrain the evaluation nodes to L1 nodes only, 
our formulation has become node classification on an ego-network. 
In the preparation of this manuscript, 
we found other similar works to tackle with node classification locally on an ego-network. 
In \cite{mcauley2012learning}, the authors developed a probabilistic model 
that leverages both topology information and node-level information to discover social circles of an ego. 
DEMON \cite{coscia2012demon} proposed to solve the centralized community detection 
of large graphs by first detecting communities locally and then combine them globally. 
It first uses Label Propagation \cite{raghavan2007near} to detect community on ego-network of all nodes
and then combine locally detected communities by thresholding their intersection. 
Our observations in the experiments and those works show that ego-network is a promising direction to go. 
Stepping back to our original problem, 
we see that AUC of PPR on L2 still has some improvements to basic heuristics. 
Although there are no good ROC points for direct application, 
the benefit brought by PPR may be useful in some extensions of the current problem. 

One possible extension of the current problem is to 
allow cooperation among direct friends or a small number of collaborating observers in the network.
Another direction is the development of privacy preserving protocols for DSN which can allow users
to help each other to improve their personal community detection without requiring excessive trust. 

It is worth to note that overlapping community detection has also drawn a lot of interest in recent years. 
Our PPR proposal is readily available for overlapping community detection. 
In essence, PPR reveals the proximity of nodes to a set of seeds. 
The observer can provide different sets of seeds so as to reveal different communities.
The detecting result can be overlapping. 

In the simulation, it seems that we used more information than topology
and people concerns about privacy of the data.
We briefly discuss the data source as follows:
1) 2-hop topology is natural under most settings, 
e.g. on many OSN's, users can check their friends' buddy lists; 
2) in the EV allocation, the observer manually label some positive nodes
(or nodes he/she believes to be positive), thus requiring no collaboration from others; 
3) in the high degree heuristic, degree information can be calculated directly on 2-hop 
topology (however the degree of L2 node is under estimated). 
Note that we can not know all node labels first and then sample random positive nodes or high degree positive nodes.
In practice, user just identifies a set of positive nodes regardless of other node properties. 
This can be implemented and the effect should be similar to random node heuristic. 

\section{Conclusion}
\label{sec:conc}

In this paper, we have studied the community detection 
problem of any given user under the constraint of 
limited topology information imposed by emerging \gls{dsn}'s. 
In particular, we consider the scenario where only 2-hop topology information is
available to the given target \gls{sns} user.
Instead of following the clustering approach taken by  most traditional works of community detection, 
our formulation yields a binary classification problem which can be 
evaluated against ground-truth data collected from real-world \gls{osn}s.
We then establish a two-stage framework for this problem
and transform our problem to the finding of a good ranking function 
of nodes in the test set.  
In particular, we adapt the notion of PPR for this purpose
and justify its applicability on community detection via the 
Ink Spilling interpretation of PPR.

Our evaluation using real-world OSN data shows that 
even with the topology as small as 2-hop 
we can ``discover''  a target user's community. 
Our proposed \gls{ppr} approach  significantly outperforms  the basic version of  \gls{pr} and
two other  heuristics believed to be commonly used by large-scale OSNs in practice. 
Our study shows that manually labeling just a few positive nodes for \gls{ppr} 
can boost the performance. 

\end{document}